\documentclass{article}
\usepackage{ijcai23}

\usepackage{times}
\usepackage{soul}
\usepackage{url}
\usepackage[hidelinks]{hyperref}
\usepackage[utf8]{inputenc}
\usepackage[small]{caption}
\usepackage{graphicx}
\usepackage{amsmath}
\usepackage{amsthm}
\usepackage{booktabs}
\usepackage{algorithm}
\usepackage{algorithmic}
\usepackage[switch]{lineno}
\usepackage{booktabs}
\usepackage{graphicx}
\usepackage[raggedrightboxes]{ragged2e}
\usepackage[table,xcdraw]{xcolor}
\usepackage{rotating}
\usepackage{amsfonts}
\usepackage{comment}
\usepackage{afterpage}
\usepackage{tabularx}
\usepackage{makecell}
\usepackage[colorinlistoftodos]{todonotes}
\usepackage{pdflscape}
\usepackage{wrapfig}

\title{Dealing Doubt: Unveiling Threat Models in Gradient Inversion Attacks under Federated Learning – A Survey and Taxonomy}

\author{
Yichuan Shi$^{1}$
\and
Olivera Kotevska$^2$\and
Viktor Reshniak $^2$\and
Abhishek Singh$^1$\And
Ramesh Raskar$^1$
\affiliations
$^1$Massachusetts Institute of Technology\\
$^2$Oak Ridge National Lab
\emails
\{kotevskao, reshniakv\}@ornl.gov,
\{yshi23, abhi24, raskar\}@media.mit.edu,
}

\begin{document}
\maketitle

\begin{abstract}
Federated Learning (FL) has emerged as a leading paradigm for decentralized, privacy preserving machine learning training. However, recent research on gradient inversion attacks (GIAs) have shown that gradient updates in FL can leak information on private training samples. While existing surveys on GIAs have focused on the honest-but-curious server threat model, there is a dearth of research categorizing attacks under the realistic and far more privacy-infringing cases of malicious servers and clients. In this paper, we present a survey and novel taxonomy of GIAs that emphasize FL threat models, particularly that of malicious servers and clients. We first formally define GIAs and contrast conventional attacks with the malicious attacker. We then summarize existing honest-but-curious attack strategies, corresponding defenses, and evaluation metrics. Critically, we dive into attacks with malicious servers and clients to highlight how they break existing FL defenses, focusing specifically on reconstruction methods, target model architectures, target data, and evaluation metrics. Lastly, we discuss open problems and future research directions.
\end{abstract}

\section{Introduction}
\label{sec:intro}
Federated learning \cite{mcmahan2017communication,bonawitz2019towards} (FL) is a machine learning framework in which multiple clients collaboratively train a shared model without exposing their private data. In the simplest configuration called FedSGD \cite{konevcny2015federated}, a central server initializes the shared model and communicates model parameters to the clients. Each client then independently trains the model on their own dataset, then sends the gradient update back to the central server. The central server aggregates these updates and adjusts the shared model's parameters, which are then sent back to the clients. To address the communication overhead of sharing gradients with every round, FedAVG is proposed to allow clients to send model parameter updates to the central server instead of gradients \cite{mcmahan2017communication}. 

Federated Learning (FL) configurations can be categorized as either \textit{horizontal}, where users possess distinct datasets sharing common features, or \textit{vertical}, where users possess different features from the same dataset. In terms of system setup, FL may involve \textit{cross-device} scenarios, wherein numerous light-weight devices -- such as IoT -- participate with limited computing power and sporadic data availability; or \textit{cross-silo} setups, where a smaller user base participates with robust computing capabilities and consistent dataset access throughout the entire training rounds.

Of course, both gradient updates and parameter updates reveal underlying information about the training data. As a result, GIAs (Gradient Inversion Attacks) have emerged as a privacy attack that seeks to reconstruct, at time $t$, client $i$'s private data and label $x_{t, i}, y_{t, i}$ using parameter information $W_{t,i}$ via the gradient $\nabla _{W_{t,i}} \mathcal{L}(W_{t,i}) = \frac{ \partial\mathcal{L}(f(x_{t, i}, W_t), y_{t, i})}{ \partial W_{t,i}}$ under FedSGD. Alternatively, reconstruction can also happen with the FedAVG algorithm \cite{mcmahan2017communication}, where $W_{t, i}$ is calculated as $W_{t, i} \xleftarrow{} W_{t, i} - \eta \nabla \mathcal{L}(W_{t, i}; b)$ for batch $b \in B$ total batches across $E$ total epochs per round.

Earlier works take inspiration from shallow leakages that naturally exist in fully connected layers \cite{zhu2019deep,zhao2020idlg} to reconstruct private data from gradient updates only. Later variants innovate on optimization or analytic methods to improve attack accuracy and scalability, but fundamentally seek to breach privacy from gradient or parameter information alone \cite{geiping2020inverting,wei2020framework,yin2021see,jin2021cafe,huang2021evaluating,chen2021understanding,jeon2021gradient,gupta2022recovering,yue2023gradient,kariyappa2022cocktail}. This is the \textit{honest-but-curious} server model, in which the server neither maliciously modifies the model nor interferes with the FL protocol, but only tries to reconstruct user data from information it can normally access. Generally, attacks under this threat model can accurately reconstruct single image data points, but struggle to scale up to larger batch and image sizes \cite{huang2021evaluating,zhang2022survey}. As a result, a natural defense against this type of attack is aggregation.

In contrast, an emergent strain of GIAs relax the assumption that participants in FL training are benign to also consider the case of \textit{malicious} servers and clients. Malicious servers could send inconsistent models to different clients \cite{pasquini2022eluding}, delay parameter updates and exploit side channels \cite{lam2021gradient}, intentionally misclassify target classes \cite{wen2022fishing}, insert malicious components in the instantiated model \cite{boenisch2023curious,fowl2021robbing,boenisch2023reconstructing,fowl2022decepticons}, and rally sybil devices \cite{boenisch2023reconstructing}. Malicious clients could try to insert backdoors into the global model via poisoning \cite{wei2023client} or become a sybil device that aids the server in attacking a victim client \cite{boenisch2023reconstructing}. Many malicious modifications aim to exploit \textit{gradient sparsity} to reveal outlier data points. Compared to the \textit{honest-but-curious} threat model, malicious attacks could scale up to arbitrarily large batch sizes and token count, break existing defenses, and repeatedly extract the same target input. 

While prior reviews have surveyed existing works on benign/passive attackers \cite{huang2021evaluating,zhang2022survey,wainakh2022federated,enthoven2020overview}, we systematically study existing works where the attacker is malicious. Honest-but-curious attackers could generally be defeated with aggregation techniques and larger batch sizes. Attacks utilizing malicious central servers and clients, however, often circumvent secure aggregation \cite{bonawitz2016practical} and other defenses. Moreover, malicious participants challenge the security and privacy assumptions in centralized federated learning -- i.e. federated learning using a central server. Thus, this paper makes the following key contributions: 

\begin{itemize}
    \item We present a novel taxonomy of GIA that focuses on threat models. We categorize attacks based on whether they assume the \textit{honest-but-curious server}, \textit{malicious server}, or \textit{malicious client}. Unlike prior taxonomies based on optimization objectives, our classification clarifies underlying threat assumptions in practical security efforts. We further categorize attacks using malicious participants based on reconstruction technique: \textit{analytical decomposition}, \textit{gradient sparsification}, or \textit{gradient isolation}. We specifically highlight attacks that have been tested against defenses and their results.
    \item For honest-but-curious server models, we enrich prior surveys by updating the catalog of attacks to include most recent publications. Importantly, we provide a contextual framework by integrating defense strategies and evaluation metrics. Formally categorizing attack success metrics, we distinguish between those rooted in model architecture, latent space, and feature space.
    \item We finally conclude the survey by discussing open problems and future research directions. Aligning with the survey's structure, we sequentially discuss emerging trends in attacks, defenses, and evaluation metrics. We specifically highlight the need to develop and evaluate attacks in a practical, realistic FL setting, as well as a case for decentralized FL.
    \item To the best of our knowledge, this is the first survey that overviews GIAs using \textit{malicious participants}. This is also the first effort to taxonomize GIAs on the basis of threat models and in context of defenses and evaluation metrics.
    
\end{itemize}

\section{Background}
\label{sec:backgr}
Even without intentional reconstruction, past works \cite{wen2022fishing,fowl2021robbing,zhu2019deep} have pointed out that linear layers with ReLU activation functions can naturally leak input information. This is because gradients in fully connected NN layers are linear combinations of their inputs. As a result, if the ReLU activation activates for only one feature out of the input, then that feature would be leaked.
\subsection{Attacks}
Generally, classic GIAs follow one of the two reconstruction methods: a) optimization based reconstruction, where the attacker minimizes a dual-optimization problem, or b) analytical reconstruction, where the attack recursively reconstructs the input from the gradient using knowledge about the model structure. Attacks are generally deemed successful if \textit{one-shot} attacks are successful, i.e., if the attacker can recover at least one image from the batch.

\paragraph{Optimization-based Reconstruction} broadly aims to minimize the distance between a randomly initialized sample gradient and the actual gradient. Also known as \textit{iterative reconstruction}, this method fundamentally solves a dual-optimization objective \cite{zhu2019deep,zhao2020idlg}
\begin{align}
x^*_i = \arg\min_{\hat{x}}\|G^t_i - \hat{G}\|^2 \\
y^*_i = \arg\min_{\hat{y}}\|G^t_i - \hat{G}\|^2
\end{align} 
where $G^t_i $ is the original gradient and $ \hat{G}$ is the dummy gradient computed from randomly instantiated data $(\hat{x}, \hat{y})$ as $\hat{G} = \nabla_{W^t} \mathcal{L}(f_{W^t}(\hat{x}), \hat{y})$. The accuracy and stability of this optimization approach can be improved with knowledge of mini-batch statistics \cite{geiping2020inverting}. Later works improve this optimization procedure with generative \cite{jeon2021gradient,fang2023gifd,hitaj2017deep} or adversarial models \cite{wu2023learning}, transformers \cite{lu2021april}, or novel data-sets \cite{hatamizadeh2023}. Additionally, \cite{jin2021cafe} applies the distance minimization technique to vertical federated learning, where each client maintains a unique features set.

Overall, optimization based approaches can attack more model architectures such as CNN and transformers. However, they tend to struggle when scaling up to larger batch sizes and may encounter issues with convergence instability from the randomly instantiated dummy images. GAN-based methods are also computationally heavy and often assume the attacker to have prior knowledge of the victim's data distribution.

\paragraph{Analytical-based Reconstruction} avoids instabilities associated with the dual optimization problem by starting from the tail-end of the model and analytically reconstructing the input to layers. While the majority of attacks under this method recursively reconstruct input layer by layer \cite{chen2021understanding,zhu2021rgap}, a recent attack \cite{kariyappa2022cocktail} frames reconstruction as a blind source separation problem (BSS) on a fully connected layer, and thus uses principle component analysis to reconstruct it. 

In contrast to optimization-based attacks, analytical attacks can better scale to large batch sizes and avoid the dual-optimization problem. However, current attacks only target fully connected and convolutional layers, and cannot target mini-batch training \cite{zhang2022survey}.
\subsection{Defenses}
Prior surveys \cite{zhang2022survey,wainakh2022federated,enthoven2020overview} broadly categorize defense mechanisms under federated learning into three classes based on the mechanism for additional privacy: cryptography, perturbation, and sanitization.
\paragraph{Cryptographic} approaches employ methods such as homomorphic encryption and multi-party computation to encrypt gradient information \cite{bonawitz2016practical,jin2023fedmlhe}. Additionally, \cite{huang2021instahide} devised an encryption scheme for training images. While recent encryption-based schemes seek to optimize performance, cryptographic defenses still pose a greater overhead on computation time and model size under a practical setting. Moreover, defenses based on key exchange cannot safeguard the FL setup against malicious clients \cite{wei2023client}.
\paragraph{Obfuscation} based methods seek to hide user data through means such as additive noise or trimming down the gradient information. Defenses on the front of noise addition include local and distributed differential privacy (DP) \cite{wei2019federated}. Measures that obfuscate via gradient reduction originate from the need to reduce network overhead in gradient and parameter exchanges, and they include methods such as gradient compression \cite{haddadpour2020federated} and pruning \cite{zhu2019deep}. While compression has not been tested against attackers, there have been attacks that recovered gradients under distributed differential privacy \cite{boenisch2023reconstructing,wu2023learning} and pruning \cite{huang2021evaluating}.

Secure aggregation \cite{bonawitz2016practical}, on the other hand, requires that clients aggregate their updates before sending it to the server. In essence, each user $u$ sends to the server an update $y_u = x_u + p_u + \sum_{v \in U_2}^{} p_{u,v}$ where $x_u$ is the output, $p$ is a pseudorandom generator function, and $v$ is a different user in the set of all users. While secure aggregation generally provides security against GIAs under the honest-but-curious server setting, later attacks using malicious participants -- detailed in Section 3 -- demonstrate that secure aggregation is not sufficient to defend against the malicious attacker either alone or even when paired with distributed differential privacy.

\paragraph{Sanitization} techniques, in contrast, generally take a more protocol-level view to root out inversion attacks. Byzantine-robust aggregation schemes (ARG) root out malicious participants by analyzing their update according to algorithms that evaluate the resulting distribution or error. Compared to other Byzantine-robust techniques such as those based on redundancy or trust, ARGs do not require trusted clients or datasets \cite{Li_2023}. However, a client side poisoning attack \cite{wei2023client} shows that, with a cleaver and stealthy attacker, such statistical margins could be manipulated to fool Byzantine-robust ARGs.

\subsection{Success Evaluation Metrics}
Evaluating the effectiveness of gradient inversion attacks requires robust and informative metrics.
In imaging domain, $L$-norms and traditional perceptual metrics (SSIM, PSNR, FSIM, etc., \cite{Wang2004image}) are widely used but often fail to capture the semantic context of recovered data, particularly in multidimensional outputs like images.
These metrics often assume pixel-wise independence, overlooking spatial relationships and crucial information within noisy and altered images.
As a result, they can miss subtle yet significant data leakage, leading to inaccurate assessments of attack success \cite{papadopoulos2023absolute}.

Emerging approaches that showed promise in capturing the semantic context and perceptual nuances are commonly feature-based.
Learned Perceptual Image Patch Similarity (LPIPS) score is the notable example of such metric between two images that is evaluated in the representation space of neural networks trained on challenging visual prediction and modeling tasks \cite{Zhang_2018_CVPR}.
Image Identifiability Precision (IIP) is a related metric that measures the amount of leaked identifiable information revealed by gradient inversion \cite{yin2021see}. 
IIP achieves this by counting the fraction of exact matches between the original image and the nearest neighbor to its reconstruction evaluated in a deep feature embedding space.
A variant of IIP computed in pixel space has been also proposed as a model-independent but more strict metric \cite{fowl2021robbing}.
Another extension of IIP was proposed in \cite{hatamizadeh2023} by computing it for the entire training and validation data of all clients, not just a random sample, offering a comprehensive picture of potential data leakage.

In an effort to address the considerable cost
of deep learning based approaches, alternative lightweight metrics utilizing linear or handcrafted features were also suggested.
Among others, frequency domain cosine similarity was used in \cite{yin2021see} to capture high-frequency details that often missed by pixel-wise comparisons.
Absolute Variation Distance (AVD) is another simplistic metric proposed in \cite{papadopoulos2023absolute}, it evaluates distances in gradient space and offers a continuous measure for information retrieval in noisy images, aligning closely with human perception.

To measure data leakage across multiple clients with different training data and privacy levels, \cite{hatamizadeh2023} proposed a Relative Data Leakage Value (RDLV) indicator based on a given similarity metric such as, e.g., SSIM.
RDLV evaluates normalized changes relative to the selected reference point enabling comparison of base metric across the clients.
To quantify latent information leakages in property inference attacks, \cite{mo2021layerwise} proposed two additional metrics: a) empirical $\mathcal{V}$-information (\cite{Xu2020A}) measures the information flow between gradients and sensitive properties (e.g., gender, race) in each layer of the FL model, and b)~Jacobian-based metric  measures the sensitivity of gradients to input changes and properties: non-sensitive gradients indicate lower attack success, providing valuable insights into the attack's vulnerability.
These metrics provide a useful insight into which clients or parts of the model are most vulnerable to privacy attacks revealing opportunities for the design of potential defenses.

While most existing GIAs focused on imaging problems, language tasks have been somewhat overlooked despite that most successful industrial applications of FL are for textual data.
Partially, this is a result of the previously accepted opinion that transformer-based models are hard to invert contributing to their privacy preservation without the need for stronger defences.
However, \cite{fowl2022decepticons} showed that malicious servers can significantly corrupt transformer models at the client side breaking their privacy.
For the attack evaluation, they used standard NLP token accuracy metrics such as Bilingual Evaluation Understudy (BLEU) and Recall-Oriented Understudy for Gisting Evaluation (ROUGE).

\section{Malicious Participants}
\label{sec:mal}

In this section, we detail our taxonomy of GIAs involving malicious servers and clients. For representative studies in each threat model, we categorize them into three sub-categories by how they extract gradient information: \textit{Analytical Decomposition}, \textit{Gradient Sparsification}, and \textit{Gradient Isolation}. We further analyze them on the basis of \textit{defenses broken}, \textit{targeted data and model}, and \textit{FL setup}. Detailed summaries and comparison of the selected studies for malicious server and client attacks are listed in Figure~\ref{tab:title}.

\subsection{Attacks}
Compared to the honest-but-curious server model, malicious participants are uniquely adept at breaking secure aggregation \cite{bonawitz2016practical} and isolating the individual gradient for a target data-point from the dense, batched aggregate, a task that is often difficult for the honest-but-curious server threat model \cite{zhang2022survey}. To pick out individual gradients from the batch, the attacker often uses side channel information \cite{lam2021gradient}, malicious model modifications \cite{fowl2022decepticons,fowl2021robbing,boenisch2023curious,pasquini2022eluding}, or malicious parameter modifications \cite{wen2022fishing,wei2023client}. 

Once the target gradient is sufficiently disaggregated, the attacker can use standard GIA techniques such as gradient matching \cite{zhu2019deep,geiping2020inverting,zhao2020idlg} to recover the original data in the case of image reconstruction. In the case of language reconstruction, the attacker needs to take additional steps to sort recovered tokens in their proper sequence. This can be achieved via a lookup dictionary that maps embeddings back to tokens \cite{boenisch2023curious} or by solving a linear sum assignment problem \cite{fowl2022decepticons}.

\renewcommand{\arraystretch}{1.5}

\begin{figure*}[t]
\resizebox{\textwidth}{!}{%
\begin{tabular}{p{2cm}|p{2.5cm}p{2.5cm}p{2.5cm}p{2.5cm}p{2.5cm}p{2.5cm}p{2.5cm}p{2.5cm}}
\hline
\footnotesize \textbf{Name} &
\textbf{\footnotesize \cite{lam2021gradient}} &
\textbf{\footnotesize \cite{boenisch2023curious}} &
\textbf{\footnotesize \cite{fowl2021robbing}} &
\textbf{\footnotesize \cite{wen2022fishing}} &
\textbf{\footnotesize \cite{fowl2022decepticons}} &
\textbf{\footnotesize \cite{wei2023client}} &
\textbf{\footnotesize \cite{pasquini2022eluding}} &
\textbf{\footnotesize \cite{boenisch2023reconstructing}} 
\\
\hline \hline

\footnotesize Threat Model &
\cellcolor[HTML]{FFE2E0}\footnotesize Adversarial Server + Side Channel &
\cellcolor[HTML]{FFCCC9}\footnotesize Malicious Server &
\cellcolor[HTML]{FFCCC9}\footnotesize Malicious Server &
\cellcolor[HTML]{FFCCC9}\footnotesize Malicious Server &
\cellcolor[HTML]{FFCCC9}\footnotesize Malicious Server &
\cellcolor[HTML]{F2D1B8}\footnotesize Malicious Client &
\cellcolor[HTML]{FFCCC9}\footnotesize Malicious Server &
\cellcolor[HTML]{FFD2C9}\footnotesize Malicious Server + Sybil Devices 
\\
\hline

\footnotesize Reconstruction Method &
\cellcolor[HTML]{C8FFFE} \footnotesize Matrix refactorization &
\cellcolor[HTML]{C8FFCE} \footnotesize Sparsity by trap weights & 
\cellcolor[HTML]{C8FFCE} \footnotesize Sparsity by imprint module &
\cellcolor[HTML]{C8FFCE} \footnotesize Sparsity by malicious parameters & 
\cellcolor[HTML]{C8FFCE} \footnotesize Sparsity by imprint module &
\cellcolor[HTML]{C8FFCE} \footnotesize Sparsity by model poisoning &
\cellcolor[HTML]{FFFCC8} \footnotesize Isolation by inconsistent model &
\cellcolor[HTML]{FFFCC8} \footnotesize Isolation by sybil devices 
\\
\hline

\footnotesize Target Data &
\footnotesize CIFAR10 and CIFAR100. \textbf{Batch sizes} 18, 16, 32. User dataset size 64, 128. &
\footnotesize MNIST, CIFAR100, ImageNet. IMDB for text. \textbf{Batch size} up to 100 &
\footnotesize ImageNet. \textbf{Batch size} varying between 64, 128, 256. & 
\footnotesize ImageNet. \textbf{Batch size} up to 250 tested. &
\footnotesize Wikitext, \textbf{Batch size} 512 w/ seq. len. 32. Stackoverflow, \textbf{batch size} 64 w/ seq. len. 512. &
\footnotesize CIFAR10, TinyImageNet, Caltech256, \textbf{Batch Size} 8-32. &
\footnotesize CIFAR10, CIFAR100, tinyImagenet. \textbf{Batch size} up to 512. &
\footnotesize CIFAR10 across mini \textbf{batch sizes} 10, 20, 100. IMDB for sentiment analysis 
 \\
\hline

\footnotesize Target Model Architecture &
\footnotesize LeNet CNN with 3 layers &
\footnotesize Networks with fully connected layers and ReLU activation - FC-NNs and CNNs & 
\footnotesize ResNet18 tested, but \textbf{Model Agnostic} for any neural network that gets non-zero gradient signal& 
\footnotesize ResNet18 with Adam optimizer tested, \textbf{Model Agnostic}& 
\footnotesize 3 layer transformer, BERT-Base, and smallest GPT2 &
\footnotesize Resnet18 but any deep neural network &
\footnotesize Resnet, but should be \textbf{model agnostic} for any neural network with fully-connected layer. &
\footnotesize 6-layer fully connected neural net, same as Curious Abandon Honesty 
 \\
\hline

\footnotesize FL Setup &
\footnotesize FedAVG. Central server is adversarial and fixes the model across rounds. Random client selection / device participation of up to 1000 users. &
\footnotesize FedAVG. Central server target users by sending model with trap weights. Users hold five mini batches ranging from $\{10, 20, 40\}$ data points. &
\footnotesize Both FedAVG and FedSGD. FedAVG case necessitates additional linear scaling in imprint module. & 
\footnotesize FedSGD, assuming attacker (central server) can estimate CDF of distribution for any feature of interest in the target class. & 
\footnotesize FedSGD with up to 100 users. Claims FedAVG &
\footnotesize FedAVG and FedSGD with byzantine robust aggregation rules (AGRs). $m$ Coordinated malicious clinets out of $N$ total clients &
\footnotesize FedSGD and FedAVG with secure aggregation. Malicious server can corrupt a fixed number of users, same threat model as secure aggregation paper. &
\footnotesize FedSGD under distributed differential privacy (DDP) and secure aggregation. $M-1$ sybil devices for $M$ sampled participants per round. 
\\
\hline

\footnotesize Input &
\footnotesize Aggregated gradients $G_{\text{aggre}}$, constraint window, constraint sums, number of users &
\footnotesize Minibatch aggregated gradients &
\footnotesize Batched gradient update after secure aggregation under cross-silo setting. &
\footnotesize Batched gradient update after secure aggregation, for both cross device and cross silo settings. &
\footnotesize Averaged gradient of multiple sequences of multiple tokens &
\footnotesize Model parameter updates from uncompromised server &
\footnotesize Batch gradient after secure aggregation &
\footnotesize Batch gradient after secure aggregation and DDP 
\\
\hline

\footnotesize Defense Broken &
\footnotesize SA &
\footnotesize None tested & 
\footnotesize SA &
\footnotesize SA & 
\footnotesize None tested &
\footnotesize Byzantine Robust AGRs &
\footnotesize SA &
\footnotesize SA + DDP
 \\
\hline

\end{tabular}%
}
\caption{Summary of attacks employing adversarial or malicious servers
}
\label{tab:title}

    \begin{tabular}{c|c|c}
        \cellcolor[HTML]{C8FFFE} Matrix Refactorization &
        \cellcolor[HTML]{C8FFCE} Gradient Sparsification &
        \cellcolor[HTML]{FFFCC8} Gradient Isolation
        \\

    \end{tabular}
\end{figure*}

\subsubsection{Means for Gradient Disambiguation}
As one of the earliest attacks that employs an adversarial central server, the Gradient Disaggregation attack \cite{lam2021gradient} disambiguates batched gradients by solving a binary factorization problem with the \textit{side-channel} knowledge of user participation across rounds. 

In the majority of later methods, the central server induces and exploits \textit{gradient sparsity} or \textit{gradient isolation} through malicious \textit{model modifications} or malicious \textit{parameter updates}. In those cases, both the federated learning and secure aggregation protocols would run as designed, but either all but one data point in the input batch return a zero gradient, or a sparse distribution is created out of a dense aggregation to recover sparse single updates.

\paragraph{Analytical Decomposition}
\cite{lam2021gradient} frames gradient disaggregation as a binary matrix factorization: 
\begin{align*}
    G_{aggregated} = P\cdot G_{individual},
\end{align*}
where $G_{aggregated} \in \mathbb{R}^{n \times d}$, $P \in \{0,1\}^{n\times u}$, and $G_{individual} \in \mathbb{R}^{u\times d}$ across $d$ dimensions, $n$ rounds, and $u$ total participants. Additional constraints are imposed using device analytics' summary information that details how often a specific user participates in training. Specifically, $C^i_k \in \{0,1\}^n$ denotes the training rounds for which total number of participants $c^i_k$ is known for the $k$th participant. Thus, with the additional constraint $C_kp_k - c_k = 0$, the user participation matrix $P$ is solved by the following optimization:
\begin{align}
\label{Lam}
& \text{Find } p_k \text{ s.t. } Nul(G^T_{aggregated})p_k = 0, \\
& p_k \in \{0, 1\}^n, \\
& C_kp_k - c_k = 0. 
\end{align}
where $Nul$ is the null space of the matrix. The final $P$ then reveals individual model updates.

\paragraph{Gradient Sparsification} 
Without explicit matrix decomposition, the compromised server could also insert malicious model modifications \cite{fowl2021robbing,fowl2022decepticons,boenisch2023curious} or parameter modifications \cite{wen2022fishing} to create gradient sparsity and isolate individual data points.

\cite{fowl2021robbing} utilizes the natural leakage in linear layers detailed in section~\ref{sec:backgr} to construct a linear imprint module with malicious weights and bias $W_*, b_*$ such that $\langle W^i_*, x\rangle$ creates a distribution of user data (for instance, image brightness) and $b^i_* = -\Phi^{-1}(\frac{i}{k}) = -c_i$, where $\Phi^{-1}$ is the inverse of the standard Gaussian CDF and $i$ denotes the $i^{th}$ row (or channel) of $W_*$ and the $i^{th}$ entry of $b_*$. The bias term creates $k$ bins according to the CDF of the distribution created by $\langle W^i_*, x\rangle$. In the example case of image brightness, if some image $x_t$ land between two distinct brightness values $c_l$ and $c_{l+1}$ such that $c_l \leq h(x_t) \leq c_{l+1}$, the imprint module can thus reconstruct
\begin{align}
\label{fishing}
& (\nabla_{W^l_*} \mathcal{L} -\nabla_{W^{l+1}_*} \mathcal{L}) \oslash \left(\frac{ \partial \mathcal{L}}{ \partial b^l_*} -\frac{ \partial \mathcal{L}}{ \partial b^{l+1}_*}\right) 
\\
& =x_t + \sum_{s=1}^{p} x_{i_s} - \sum_{s=1}^{p} x_{i_s} 
\\ \nonumber
 & =x_t,
\end{align}
where every $x_i$ is a image from the batch that has brightness greater than $c_l$.

Similarly, \cite{fowl2022decepticons} extends this idea of measuring and then capturing induced sparsity to language transformers. Malicious modifications to the transformer first disable all attention layers and most outputs of each feed-forward block, such that only the last entry has a non-zero gradient and token information is not mixed. The reconstruction task then only deals with the gradients of the first linear layer in each feed-forward block. The operation of the first linear layer in each $i^{th}$ feed-forward block is $y_k=W_i u_k + b_i$, where $\{u_k\}^S_{k=1}$ represents the input embedding that the attacker wants to recover, in a sequence of $S$ elements. The malicious $\langle W_{*_i}, b_{*_i} \rangle$ in \cite{fowl2021robbing} are then used for each linear layer $i$ to separate gradient signals. Crucially, the bias terms over all layers are set to be \textit{sequentially ascending}, so the reconstructed $y^j_k = \langle W_*, u_k\rangle + b^j$ have the ordering $y^j_k \geq y^{j+1}_k, \forall j < j+1$. When subsequent ReLU units are thresholding at $y_k^j$, embedding ordering can be uncovered by the following subtractions:
\begin{align}
\sum^S_{k=1} \nabla_{W^j_i} \mathcal{L}_k - \sum^S_{k=1} \nabla_{W^{j+1}_i} \mathcal{L}_k & = \lambda^j_i u_k, \\
\sum^S_{k=1} \frac{\partial \mathcal{L}_k}{\partial b^j} - \sum^S_{k=1} \frac{\partial \mathcal{L}_k}{\partial b^{j+1}} & =\lambda^j_i.
\end{align}

In addition, multiple sequences are disambiguated with query, key, and value weight matrices and biases $\langle W_Q, b_Q\rangle, \langle W_K, b_K\rangle, \langle W_V, b_V\rangle$. The query, key, and value matrices are initialized such that $W_K = I_d, b_K = \overrightarrow{0}, W_Q=\overrightarrow{0},b_Q=\gamma \overrightarrow{p}_0, W_V=I_{d'}, b_V=\overrightarrow{0}$ where $\overrightarrow{p}_0$ is the first positional encoding, and $\gamma, d'$ are server hyperparameters. After passing attention weights through the softmax function $\sigma(\frac{QK^T}{\sqrt{d_k}})$, the recovered embeddings from each linear layer contain unique information about each sequence's first tokens. Sequence disambiguation can then be performed by calculating correlations between such first-position embeddings and later-position embeddings via clustering algorithms such as K-means.

Building off the idea of sparsification via model modification, but recognizing that the imprint layer design could be more easily detectable by the client, \cite{boenisch2023curious} instead constructs "trap weights". Recognizing and amplifying the previous discussed natural gradient leakage in fully-connected linear layers, the "Curious Abandon Honesty" attack clarifies that under ReLU activation, a neuron $i$ is only active if the sum of features weighted by the negative weights $w^n_i$ is less than the sum of weights that are positive $w^p_i$: 
\begin{align}\label{eq:myeq}
\sum^{}_{n \in N}w^n_i x_n < \sum^{}_{n \in N}w^p_i x_p   
\end{align}
Thus, the adversarial server initialize the model such that roughly half of the row weights are negative, and the other half positive. By controlling the probability that Equation~\eqref{eq:myeq} holds, the adversarial initialization activates for only one input data point per mini-batch, thereby disaggregating batched gradients.

Lastly, \cite{wen2022fishing} takes the idea of stealth one step further and constructs a server-side privacy attack that only uses compromised parameters, rather than any detectable model modifications. The attack builds off the crucial observation that when the confidence of a target class' prediction \textit{decreases}, its corresponding gradient contribution to the gradient information \textit{increases}. As a result, the \textit{class fishing} attack modifies a fully-connected classification layer $\langle W \in \mathcal{R}^{n \times m}, b \in \mathcal{R}^n \rangle$ to isolate target class $c$ as follows:
\begin{align}
W_{i,j}=
\begin{cases}
W_{i,j} & \text{ if } i=c, \\
0 & \text{ otherwise},
\end{cases}
\end{align}
\begin{align}
b_{i}=
\begin{cases}
b_i & \text{ if } i=c, \\
\alpha & \text{ otherwise}.
\end{cases}
\end{align}
where $\alpha$ is a server-side hyperparameter. To address intra-batch class collisions, a separate \textit{feature-fishing } strategy is devised using the same idea. Both $\theta$ and $\beta$ are server-side hyperparameters.
\begin{align}
W_{i,j}=
\begin{cases}
\beta & \text{ if } i=c, j=k,\\
0 & \text{ otherwise},
\end{cases}
\end{align}
\begin{align}
b_{i}=
\begin{cases}
-\beta \theta & \text{ if } i=c, \\
\alpha & \text{ otherwise}.
\end{cases}
\end{align}

Though malicious server attacks are highly effective at disaggregating gradients under protections such as secure aggregation and differential privacy, \textit{gaining control of the central server may be unrealistic}, as it is typically heavily safe-guarded. \cite{wei2023client} instead argues that a malicious client is a more realistic threat model, and conceives of the client-side poisoning Gradient Inversion (CGI) attack. The malicious client first crafts a local model that intentionally maximizes the loss of the targeted sample class by misclassification or omission of the target class in local training. Once the malicious local parameters $\Phi_{mal}$ are obtained, the client then poison the global model while evading byzantine secure aggregation rules by minimizing the distance between the malicious update $\delta_{mal}$ and the current aggregated update $\delta_{agg}$:
\begin{align}
    \delta_{poi} &= \Phi^t_G - \Phi_{mal},
    \\
    \delta_{agg} &= \frac{1}{\alpha}(\Phi^{t-1}_G-\Phi'_G),
    \\
    \delta_{mal} &= \rho \frac{\delta_{poi}}{\|\delta_{poi}\|} + \beta \hat{\delta}_{benigh},
\end{align}
where $\Phi_G$ denotes the global parameters, and $\alpha$, $\rho$, $\beta$ are tunable parameters. And thus, the adversary seeks to minimize 
\begin{align}
    \arg\min_{\gamma, \beta}\|\delta_{mal} - \delta_{agg}\|_2.
\end{align}
With the global model poisoned, the adversary can then reconstruct the targeted gradient by first calculating the aggregate update, instantiating a target gradient with randomly generated data, and reconstructing the target image by minimizing cosine distance in the same way as \cite{zhu2019deep,geiping2020inverting}. Specifically, the numerical reconstruction method is used from \cite{geiping2020inverting}:
\begin{align}
    \arg\min_{\hat{x}_{tar}} \space \space \lambda_0(1-\frac{\langle \hat{\delta}_{tar}, \delta^k_{agg}\rangle}{\|\hat{\delta}_{tar}\|\|\delta^k_{agg}\|}) & \nonumber \\ 
    + \lambda_1 R_{TV}(\hat{x}_{tar}) + & \lambda_2 R_{l2}(\hat{x}_{tar}) + \lambda_3 R_{clip}(\hat{x}_{tar}) 
\end{align}
with 
\begin{align}
    \delta^k_{agg} & = \frac{1}{\alpha}(\Phi^{k+1}_G - \Phi^k_G), \\
    \hat{\delta}_{target} & = \nabla_{\Phi^k_G}\mathcal{L}(\hat{x}_{target}, y_{target}).
\end{align}

\paragraph{Gradient Isolation}
In contract to attacks that create a sparse distribution from a dense, batched gradient, another variant of the malicious server attack instead pick out single input gradients from the aggregation by isolating the target gradient either through sending users inconsistent models \cite{pasquini2022eluding} or rallying sybil devices \cite{bonawitz2016practical}.

\cite{pasquini2022eluding} pointed out that secure aggregation provides privacy protection by performing a weighted sum across all users' gradient contributions. As a result, the malicious server breaks secure aggregation by making sure that only the victim user returns a non-zero gradient to secure aggregation. Using ReLU's ability to cancel out values less than or equal to zero, a sample modification would be to set $W = [0]$ and $b = [\mathcal{R}_{\leq 0}]$ for non-target users in the batch. For networks not using ReLU, the malicious server can insert a ReLU layer.

Building off the exploitation of zeroing other gradients' contribution to secure aggregation, \cite{boenisch2023reconstructing} conceives of a gradient-isolation mechanism that depends on a malicious central server that can rally sybil devices. This scheme not only breaks secure aggregation, but also distributed differential privacy (DDP). Given an FL protocol that requires $M$ clients to be sampled per round, the malicious server targets a single victim clients and rallies $M-1$ sybil devices. Thus, when the sybil devices contribute known, or empty, gradients to the server, secure aggregation is broken and the target user's gradient is exposed. The authors make the additional observation that the attack also breaks DDP, as its sampled local Guassian noise is dependent on the number of participants:
\begin{align*}
    \mathcal{N}\left(0, \frac{\sigma^2}{M-1}c^2\right).
\end{align*}
When the number of participants $M$ increases, the amount of noise needed from each participant proportionally decreases. As a result, if $M$ is sufficiently large and the $M-1$ sybil devices introduce less noise than necessary, or none at all, the differential privacy safeguard for the target user could significantly diminish, approaching a negligible level. 

\subsection{Performance Evaluation}
Generally, attacks under the honest-but-curious server model struggles to scale up to larger batches, higher resolution images, different data and model types such as language, and many require unlikely assumptions like access to minibatch statistics \cite{huang2021evaluating,zhang2022survey,geiping2020inverting}. On the other hand, attacks using malicious participants are generally much more successful in their scalability, reproducibility, and extensibility to tasks beyond image classification.
\paragraph{Target Data}
Classic GIAs using the honest-but-curious server generally work on simpler image datasets with smaller batches, with \cite{huang2021evaluating} having the biggest success in recovering ImageNet data with dimension of $224 \times 224$ and batch size of 100. On the language front, \cite{gupta2022recovering} achieved a ROUGUE-L score of around 0.3 for batch sizes greater than 16 when using a language model prior and beam search to aid in attack success.

On the contrary, many attacks using malicious participants can scale up to batch sizes of 100 with ImageNet data \cite{boenisch2023curious,fowl2021robbing,wen2022fishing,fowl2022decepticons,pasquini2022eluding,boenisch2023reconstructing}. Moreover, the attacks targeting language models\cite{boenisch2023curious,fowl2022decepticons} outperform comparable honest-but-curious attacks by the amount of recoverable tokens per sentence, and sentence per batch (250 tokens in \cite{boenisch2023curious}, and up to 512 sequences per batch in \cite{fowl2022decepticons}).
\paragraph{Data Distribution}
All malicious attacks are experimented on independent and identically distributed (IID) data. Though there exists works that address the challenge of data heterogeneity in a federated setting, we have not found any research pertaining to malicious-participant GIAs that involve non-IID data. On the other hand, while the majority of GIAs look at attacks in the cross-silo setting in federated learning \cite{boenisch2023curious,boenisch2023reconstructing,fowl2021robbing,pasquini2022eluding}, \citeauthor{wen2022fishing} look at both cross-device and cross-silo scenarios. 
\paragraph{FL Setup}
In classic GIAs that operate under the honest-but-curious server threat model, proposed gradient reconstruction methods often work only for the simple FedSGD setting and struggle to adapt to FedAVG, which contains epoch updates in the gradient information \cite{zhang2022survey,huang2021evaluating}. In contrast, many malicious participant GIAs successfully break FedAVG \cite{boenisch2023curious,fowl2021robbing,pasquini2022eluding,boenisch2023reconstructing,lam2021gradient}. In fact, \cite{lam2021gradient} scaled the attack up to 1000 participating users that randomly enter and exit the FL protocol. Their work demonstrates that under the malicious participant threat model, participation of any given client in the aggregation step across training epochs can be used to "disaggregate" their updates from the aggregated model. Hence, aggregation does not necessarily provide stronger security protections.

\paragraph{Defenses Broken}
As previously mentioned, the unique challenge posed by the malicious participant threat model is the ease by which they break secure aggregation and scale up to larger batch sizes. Besides secure aggregation, however, only distributed differential privacy has been broken by the attack proposed in \cite{boenisch2023reconstructing}.

\section{Future Research}
\label{sec:fut}

\subsection{Improving Evaluation}
Evaluating reconstruction attack is difficult for high dimensional data because the intrinsic dimension of data is smaller as it lies on a manifold. For modalities like images, different evaluation metrics come with their own trade-offs and hence, quantifying attack success solely based on a scalar needs to account for practical considerations. Domain-specific metrics, like comparing embeddings from large language models for text is perhaps the most effective approach to capture semantic similarity.

Many times pixel perfect reconstruction is not a practical requirement anyway because attackers might be interested in inferring sensitive information such as a sensitive attribute in an image (such as identity, race, age etc.).

Differential privacy offers a solution to the problem of various trade-offs in evaluation metrics by quantifying privacy budget in terms of $\epsilon$ (and $\delta$), however, it rests on a very specific attack model (membership inference attack) and hence may not be amenable for reconstruction attacks~\cite{chatzikokolakis2013broadening,kifer2014pufferfish,singh2023posthoc}.
\subsection{Improving Attacks}
In a similar vein of standardization, we see that an emerging trend in offensive GIA research is attacks that are viable under more realistic FL setups, including defenses and real-world datasets. For instance, \cite{hatamizadeh2023} tested the proposed GIA attack against varied differential privacy measures, clients with different amounts of data and batch sizes, and Covid-19 chest X-Ray images. Similarly, \cite{wu2023learning} benchmarked and proved the success of an adaptive attack against gradient perturbation (differential privacy), gradient pruning, and sign compression on both language and image datasets. Going forward, we anticipate attacks under all threat models to be tested against realistic defenses and FL scenarios where clients potentially dropout and rejoin at will.
\subsection{Improving Defenses}
When considering potential defense strategies, one significant scenario involves fortifying against \textit{gradient sparsification} attacks. As illustrated in Table~\ref{tab:title}, the previously employed secure aggregation defense mechanisms fell victim to breaches detailed by \cite{boenisch2023curious,fowl2021robbing,wen2022fishing,fowl2022decepticons,wei2023client}. 

One way to defend against mechanisms that intentionally sparsify gradients is to make the gradients more condensed. Techniques rooted in in data compaction to improve communication efficiency, such as sketching techniques \cite{hu2021make}, may reduce the transmission and exposure of outlier points. 

Another potential strategy involves the implementation of model shuffling techniques \cite{girgis2021shuffled}. This method compresses updates and dynamically selects a client for communication. Even if an adversary compromises a client, there remains a probability that the compromised client may not be chosen, adding a layer of resilience against attacks. Furthermore, the development of client trust strategies based on the quality of data and model parameters proves to be a promising avenue for defense. 

Lastly, while not directly a mitigation strategy, \textit{decentralized FL} may inadvertently provide defenses against the malicious threat model. Decentralized FL may employ a peer-to-peer network for transmitting, aggregating, and updating local models. Participants solely engage with their immediate neighbors in the network. Notably, this architecture poses a formidable challenge for attackers due to the absence of a centralized node communicating with all clients, rendering it a promising defense strategy for further investigation and implementation.


\section{Acknowledgement}

This manuscript has been co-authored by UT-Battelle, LLC, under contract DE-AC05-00OR22725 with the US Department of Energy (DOE). The US government retains and the publisher, by accepting the article for publication, acknowledges that the US government retains a nonexclusive, paid-up, irrevocable, worldwide license to publish or reproduce the published form of this manuscript, or allow others to do so, for US government purposes. DOE will provide public access to these results of federally sponsored research in accordance with the DOE Public Access Plan (http://energy.gov/downloads/doe-public-access-plan).

\clearpage
\bibliographystyle{named}
\bibliography{main}

\end{document}